\documentclass[a4paper]{jpconf}
\usepackage{graphicx}
\usepackage{amsmath}

\newcommand{\beq}{\begin{equation}}
\newcommand{\eeq}{\end{equation}}
\newcommand{\beqa}{\begin{eqnarray}}
\newcommand{\eeqa}{\end{eqnarray}}
\newcommand{\bseq}{\begin{subequations}}
\newcommand{\eseq}{\end{subequations}}

\def\x{{\boldsymbol x}}

\def\p{{\boldsymbol p}}
\def\0{{\boldsymbol 0}}

\def\kk{{\kappa}}
\def\pp{{\hat{p}}}

\def\cal{\mathcal}

\begin{document}
\title{Heavy-quark Langevin dynamics and single-electron spectra in nucleus-nucleus collision}

\author{W M Alberico$^{3,4}$, A Beraudo$^{1,2,3,4}$, A De Pace$^4$, A Molinari$^{3,4}$, M Monteno$^4$, M Nardi$^4$ and F Prino$^4$}

\address{$^1$Centro Studi e Ricerche \emph{Enrico Fermi}, Piazza del Viminale 1, Roma, ITALY}
\address{$^2$Physics Department, Theory Unit, CERN, CH-1211 Gen\`eve 23, Switzerland}
\address{$^3$Dipartimento di Fisica Teorica dell'Universit\`a di Torino, via P.Giuria 1, I-10125 Torino, Italy}
\address{$^4$Istituto Nazionale di Fisica Nucleare, Sezione di Torino, via P.Giuria 1, I-10125 Torino, Italy}

\ead{beraudo@to.infn.it}

\begin{abstract}
The stochastic dynamics of heavy quarks in the fireball produced in heavy-ion collisions is followed through numerical simulations based on the Langevin equation. The modification of the final $p_T$ spectra ($R_{AA}$) of $c$ and $b$ quarks, hadrons and single-electrons with respect to $pp$ collisions is studied. The transport coefficients are evaluated treating separately the contribution of soft and hard collisions. The initial heavy-quark spectra are generated according to NLO-pQCD, accounting for nuclear effects through recent nPDFs. The evolution of the medium is obtained from the output of two hydro-codes (ideal and viscous). The heavy-quark fragmentation into hadrons and their final semileptonic decays are implemented according to up to date experimental data. A comparison with RHIC data for non-photonic electron spectra is given.
\end{abstract}

\section{Introduction}
In a previous paper~\cite{lange} we evaluated the transport coefficients for heavy quarks in the QGP in the Hard Thermal Loop (HTL) approximation. The above coefficients were then used to solve the relativistic Langevin equation in the case of a static medium, following the approach to thermal equilibrium. Here we improve our previous study in two important aspects. We provide a more accurate microscopic calculation of the transport coefficients and we perform the Langevin simulations for a fully realistic scenario, accounting for the hydrodynamical evolution of the medium. For independent studies based on the Langevin approach see~\cite{hira,rapp,tea,aic}.
\section{The Langevin equation in a dynamical medium}
The relativistic Langevin equation allows to study the stochastic dynamics of heavy quarks in the QGP.
The algorithm to follow the evolution of momentum and position of the brownian particle is the following. We focus on a given quark at $(\x_n,\p_n)$ after $n$ steps of evolution. We move to the local fluid rest-frame and update its position and momentum by the quantities $\Delta\bar{\x}_n\!=\!(\bar{\p}_n/\bar E_p)\Delta\bar t$ and 
\beq
\Delta\bar{p}_n^i=-\eta_D(\bar{p}_n)\bar{p}_n^i\Delta\bar t+\xi^i(\bar t)\Delta\bar{t}
\equiv-\eta_D(\bar p_n)\bar p_n^i\Delta \bar t+{g^{ij}(\bar\p_n)}{\eta^j(\bar t)}\sqrt{\Delta\bar t},
\eeq
where we take -- in the fluid rest-frame -- $\Delta\bar t\!=\!0.02$ fm/c. In the above we express the noise term through the tensor (we omit the ``bar'')
\beq
{g^{ij}(\p)\!\equiv\!\sqrt{\kk_L(p)}\pp^i\pp^j+\sqrt{\kk_T(p)}
(\delta^{ij}-\pp^i\pp^j)},
\eeq
depending on the transverse/longitudinal momentum diffusion coefficients $\kappa_{T/L}(p)$, and the uncorrelated random variables $\eta^j$, with $\langle\eta^i(t)\eta^j(t')\rangle\!=\!\delta^{ij}\delta_{tt'}$.
Hence, one simply needs to extract three random numbers $\eta^j$ from a gaussian distribution with $\sigma\!=\!1$. Concerning the problem of fixing the friction term $\eta_D(p)$ and further details we refer the reader to Ref.~\cite{lange}. We then go back to the Lab frame, getting the updated $(\x_{n+1},\p_{n+1})$. The four-velocity and temperature fields $u^\mu(x)$ and $T(x)$ of the background medium are obtained from the output of two different hydrodynamical codes~\cite{kolb1,rom1,rom2}.
\section{Evaluation of the transport coefficients}
The coefficients $\kappa_{T/L}(p)$ reflect the transverse/longitudinal squared-momentum acquired through the collisions in the medium. Following~\cite{pei} we introduce an intermediate cutoff $|t|^*\!\sim\!m_D^2$ ($t\!\equiv\!(P'\!-\!P)^2\!=\!\omega^2\!-\!q^2$) separating hard and soft scatterings. The contribution of hard collisions ($|t|\!>\!|t|^*$) is evaluated in pQCD, from the diagrams $Qq(\bar q)\!\to\! Qq(\bar q)$ and
$Qg\!\to\! Qg$
and reads (employing the notation $\int_k\equiv \int d\vec{k}/(2\pi)^3$ and $P_{\rm tot}\!\equiv\!P\!+\!K\!-\!P'\!-\!K'$)
\beq
\kappa_T^{\rm (g/q)hard}(p)=\frac{1}{2}\frac{1}{2E}\!\!\int_k\frac{n_{B/F}(k)}{2k} \!\!\int_{k'}\frac{1\pm n_{B/F}(k')}{2k'}\!\!\int_{p'}\frac{1}{2E'}\theta(|t|-|t|^*)
(2\pi)^4\delta^{(4)}(P_{\rm tot})\left|\overline{{\cal M}}_{g/q}\right|^2 q_T^2
\eeq
and
\beq
\kappa_L^{\rm (g/q)hard}(p)=\frac{1}{2E}\!\!\int_k\frac{n_{B/F}(k)}{2k}\!\! \int_{k'}\frac{1\pm n_{B/F}(k')}{2k'}\!\!\int_{p'}\frac{1}{2E'}\theta(|t|-|t|^*)
(2\pi)^4\delta^{(4)}(P_{\rm tot})\left|\overline{{\cal M}}_{g/q}\right|^2 q_L^2,
\eeq
where the squared amplitudes were evaluated in Ref.~\cite{com}. On the other hand in soft collisions ($|t|\!<\!|t|^*$) the exchanged gluon has a small virtuality and has thus ``time'' to feel the presence of the other particles. A resummation of medium effects is required and this is provided by the HTL approximation, leading to the compact formulas ($x\!\equiv\!\omega/q$)
\beq
\kappa_T^{\rm soft}(p)=\frac{C_F g^2}{8\pi^2 v}\int_0^{|t|^*}\!\!d|t|\int_{0}^{v}dx\frac{|t|^{3/2}}{2(1-x^2)^{5/2}}\,\overline{\rho}(|t|,x)\left(1-\frac{x^2}{v^2}\right)\coth\left(\frac{x\sqrt{\frac{|t|}{1-x^2}}}{2T}\right)
\eeq
and
\beq
\kappa_L^{\rm soft}(p)=\frac{C_F g^2}{4\pi^2 v}\int_0^{|t|^*}\!\!d|t|\int_{0}^{v}dx\frac{|t|^{3/2}}{2(1-x^2)^{5/2}}\,\overline{\rho}(|t|,x)\,\frac{x^2}{v^2}\coth\left(\frac{x\sqrt{\frac{|t|}{1-x^2}}}{2T}\right),
\eeq
expressed in terms of the resummed gluon spectral function $\overline{\rho}(|t|,x)\!\equiv\!\rho_L(|t|,x)\!+\!(v^2\!-\!x^2)\rho_T(|t|,x)$ which can be found in~\cite{lange}.
\begin{figure}
\begin{center}
\includegraphics[clip,width=0.46\textwidth]{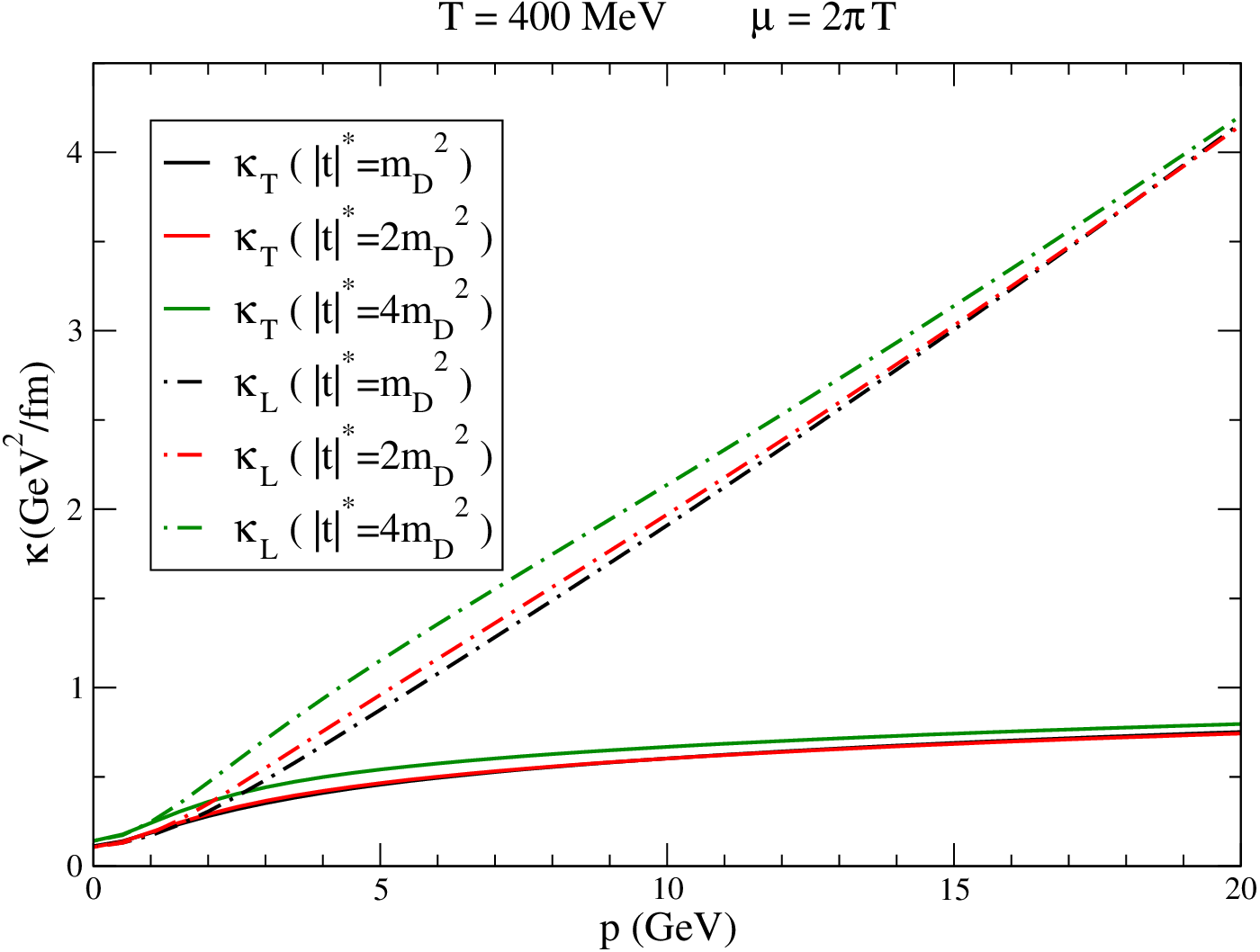}
\includegraphics[clip,width=0.49\textwidth]{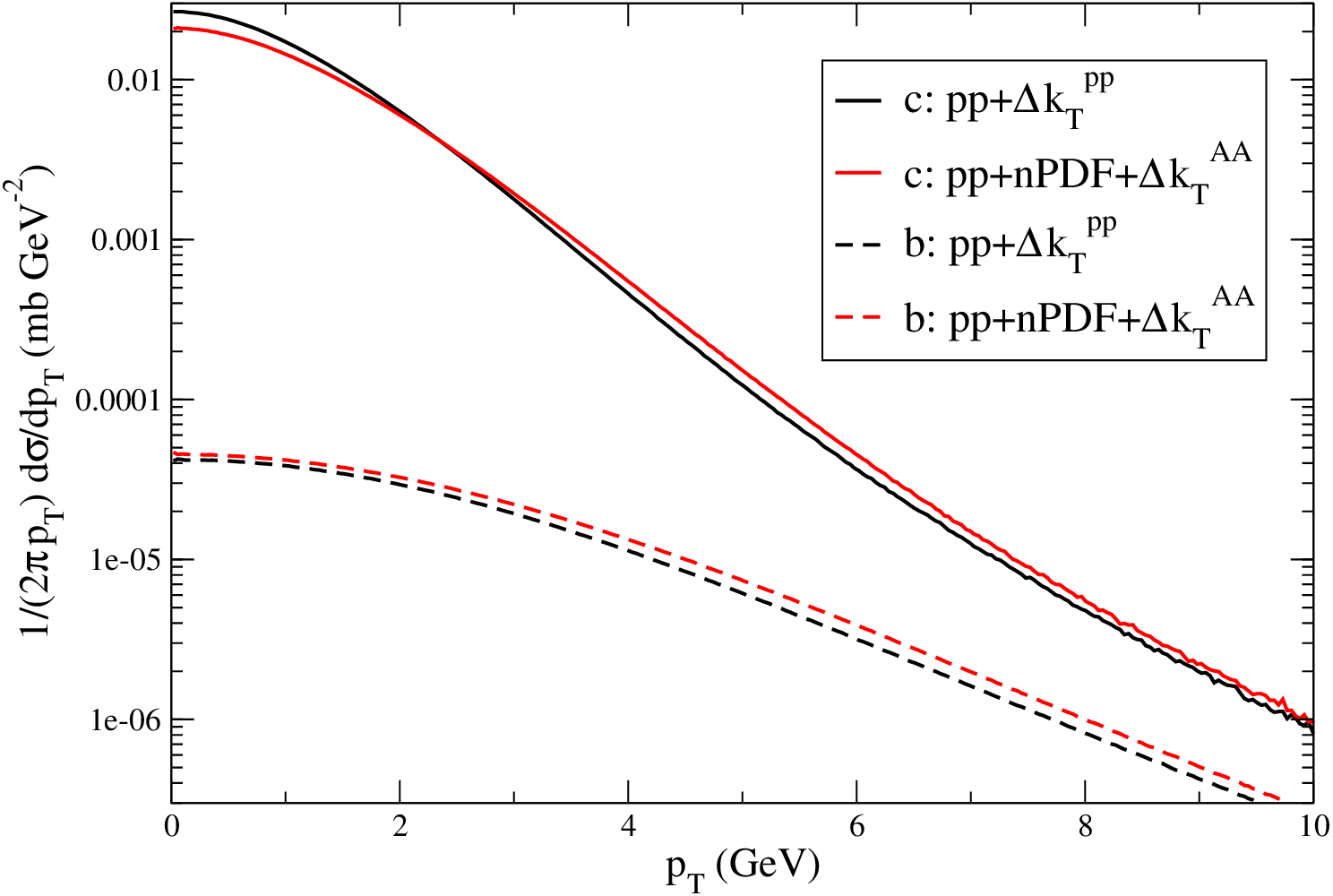}
\caption{Left panel: the momentum-diffusion coefficients $\kappa_{T/L}(p)$ after summing the soft and hard contributions. The dependence on the intermediate cutoff $|t|^*\!\sim\!m_D^2$ is very mild. Right panel: the initial $c$ and $b$ spectra generated by POWHEG for $pp$ and $AA$ (with EPS09 nPDFs and nuclear $k_T$-broadening) collisions at $\sqrt{s}_{NN}\!=\!200$ GeV.} 
\label{fig:transport}
\end{center}
\end{figure}
Our results for $\kappa_{T/L}(p)$ are shown in the left panel of Fig.~\ref{fig:transport} where a very mild dependence on the arbitrary intermediate cutoff is manifest. For large momenta $\kappa_L$ largely exceeds $\kappa_T$.
\section{Numerical results}
For each explored case we generated an initial sample of $45\!\cdot\!10^6$ $c\bar c$ and $b\bar b$ pairs, using the POWHEG code~\cite{POWHEG}, with CTEQ6M PDFs. In the $AA$ case we introduced nuclear effects in the PDFs according to the EPS09 scheme~\cite{EPS09}; the quarks were then distributed in the transverse plane according to the nuclear overlap function ${dN/d\x_\perp\!\sim\!T_{AB}(x,y)}\!\equiv\!T_A(x\!+\!b/2,y)T_B(x\!-\!b/2,y)$ and given a further $k_T$ broadening on top of the ``intrinsic'' one.
The initial $p_T$ spectra are displayed in the right panel of Fig.~\ref{fig:transport}.
For each quark, after the initial free-streaming, at the proper-time $\tau\!\equiv{t^2\!-\!z^2}\!=\!\tau_0$ we started following its Langevin dynamics until hadronization. The latter was modeled using a Peterson fragmentation function~\cite{peter}, setting $\epsilon\!=\!0.04$ and $0.005$ for $c$ and $b$ respectively; the hadron species were then assigned according to the branching fractions taken from~\cite{zeus,pdg}.
Finally each hadron was forced to decay into electrons with PYTHIA~\cite{Pythia}, using updated decay tables~\cite{pdg09}.
The $e$-spectra from $c$ and $b$ were then combined with a weight given by the respective total production cross-section ($\sigma_{c\bar c/b\bar b}$).
%

\begin{figure}
\begin{center}
\includegraphics[clip,width=0.49\textwidth]{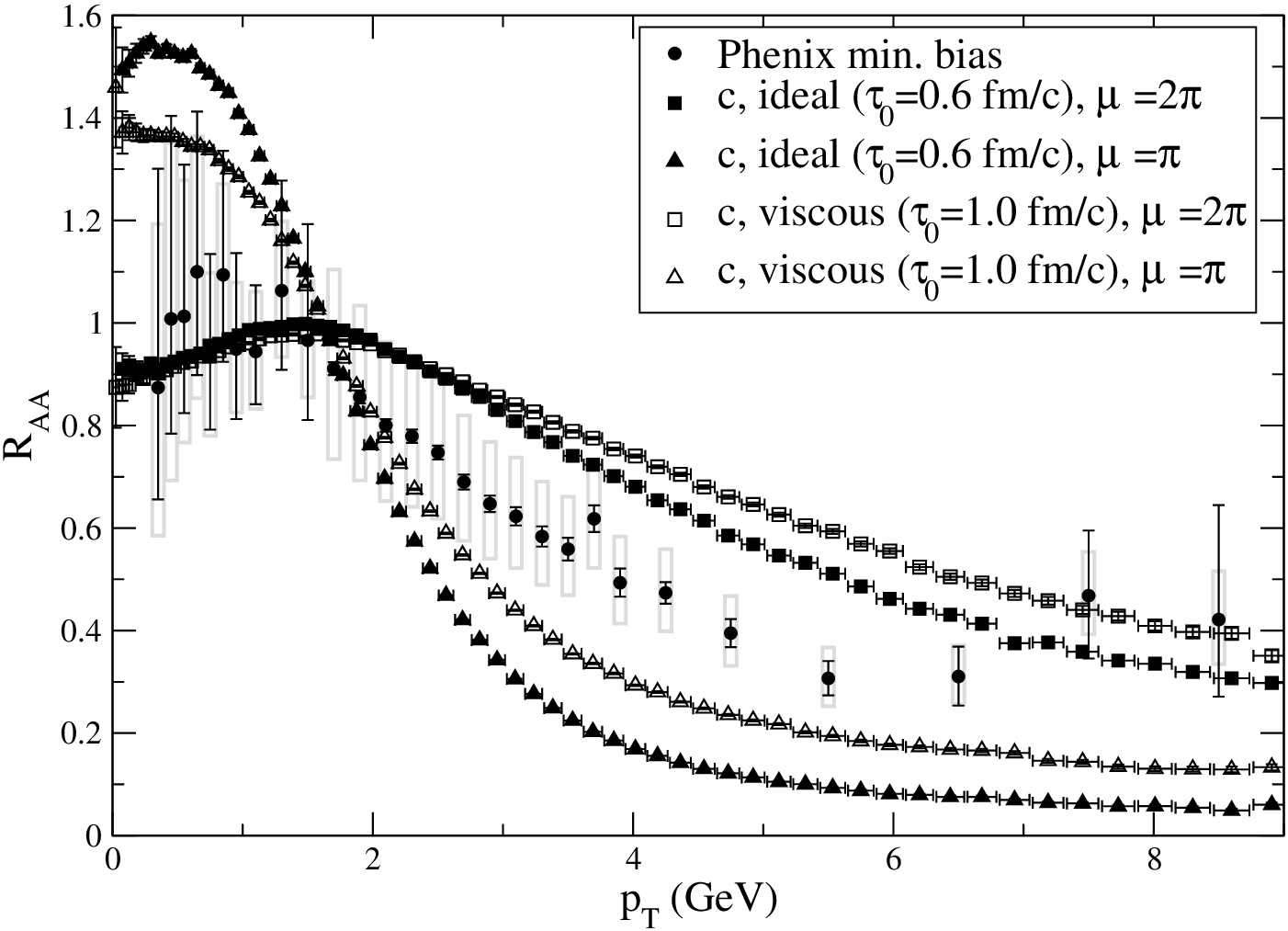}
\includegraphics[clip,width=0.49\textwidth]{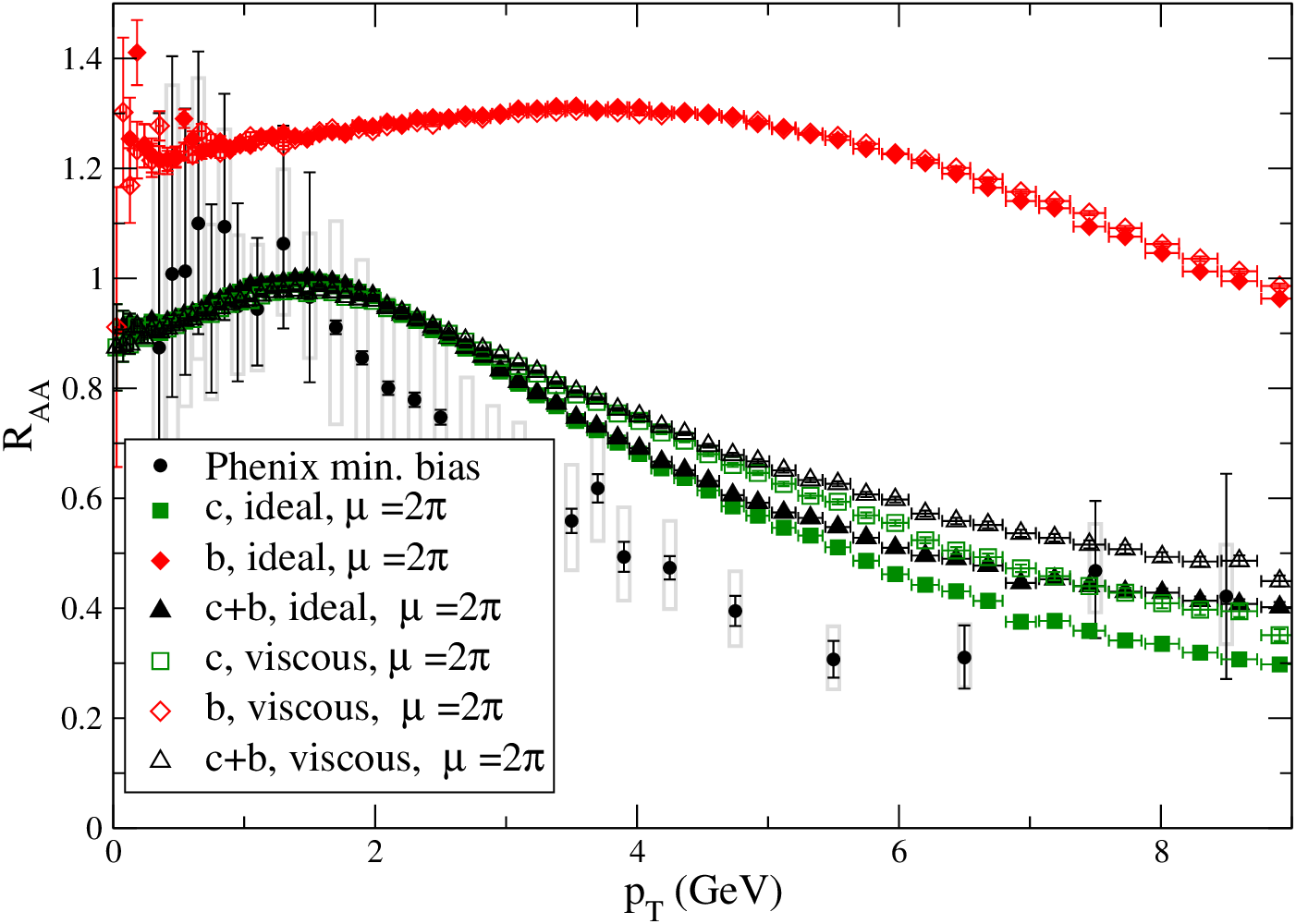}
\caption{Left panel: $R_{AA}(p_T)$ for the $c$-quark spectra after the Langevin evolution in the QGP. The sensitivity on the hydrodynamical scenario (ideal vs viscous) appears to be small. On the other hand the dependence on the coupling $g(\mu)$ (at $T\!=\!200$ MeV $\alpha_s\!\approx\!0.6/0.34$ for $\mu\!=\!\pi T/2\pi T$) is huge. Right panel: $R_{AA}(p_T)$ for $c$, and $b$ quark spectra, together with their weighted combination.} 
\label{fig:quarks}
\end{center}
\end{figure}
The results we present refer to minimum-bias Au-Au collisions at $\sqrt{s}_{NN}\!=\!200$ GeV.
In Fig.~\ref{fig:quarks} we display the $R_{AA}(p_T)\!\equiv\!(dN/dp_T)^{AA}/\langle N_{\rm coll}\rangle(dN/dp_T)^{pp}$ for the final quark spectra. The dependence on the hydro scenario appears to be small. On the other hand the results display a huge sensitivity to the value of the coupling (hence, of the transport coefficients): a too strong coupling would overestimate the quenching effect. For the $b$ spectra, the quenching induced by the medium is small: the most relevant effect is the larger production cross section in $AA$ with respect to $pp$, arising from the anti-shadowing region in the nPDFs.
\begin{figure}
\begin{center}
\includegraphics[clip,width=0.49\textwidth]{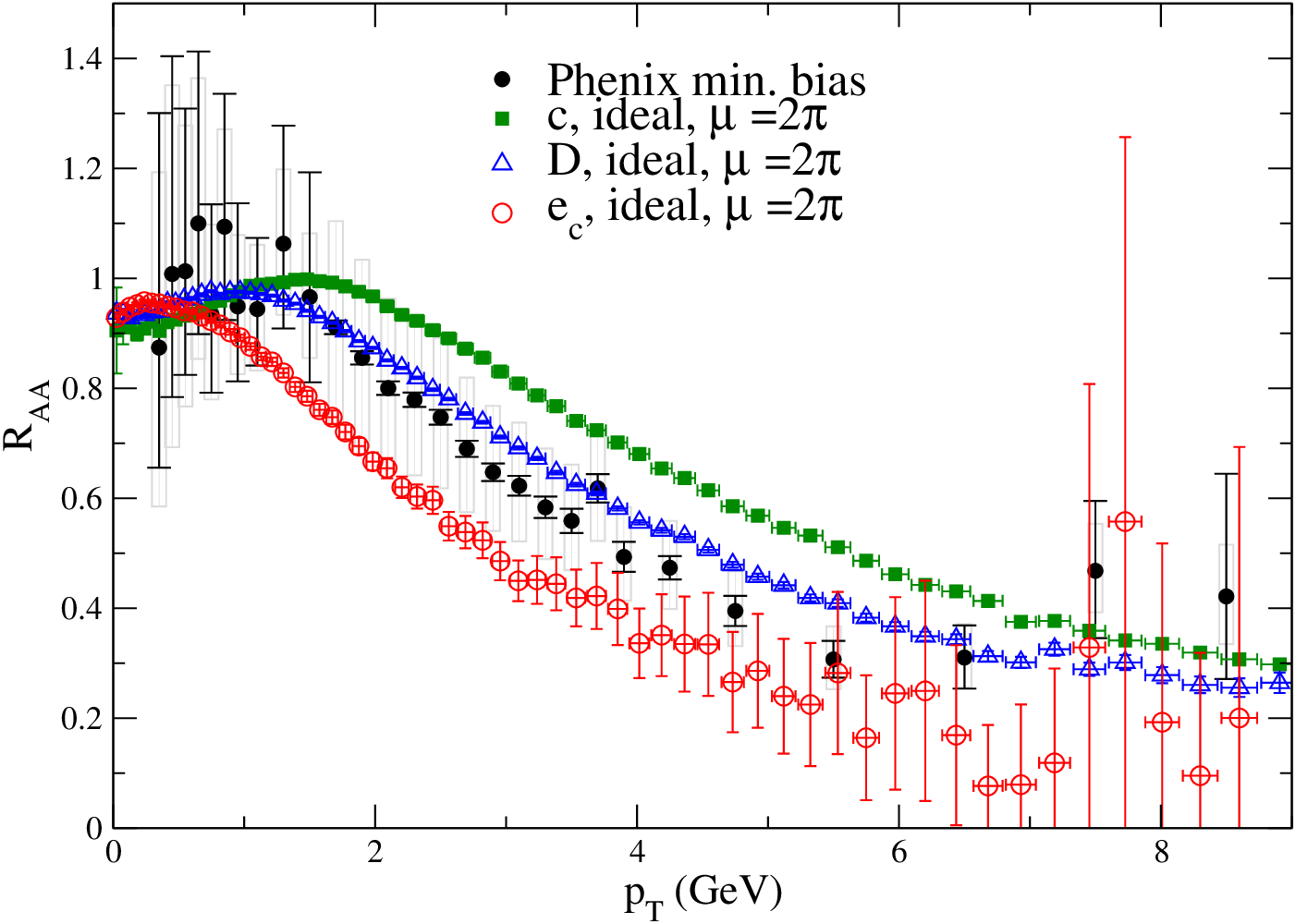}
\includegraphics[clip,width=0.49\textwidth]{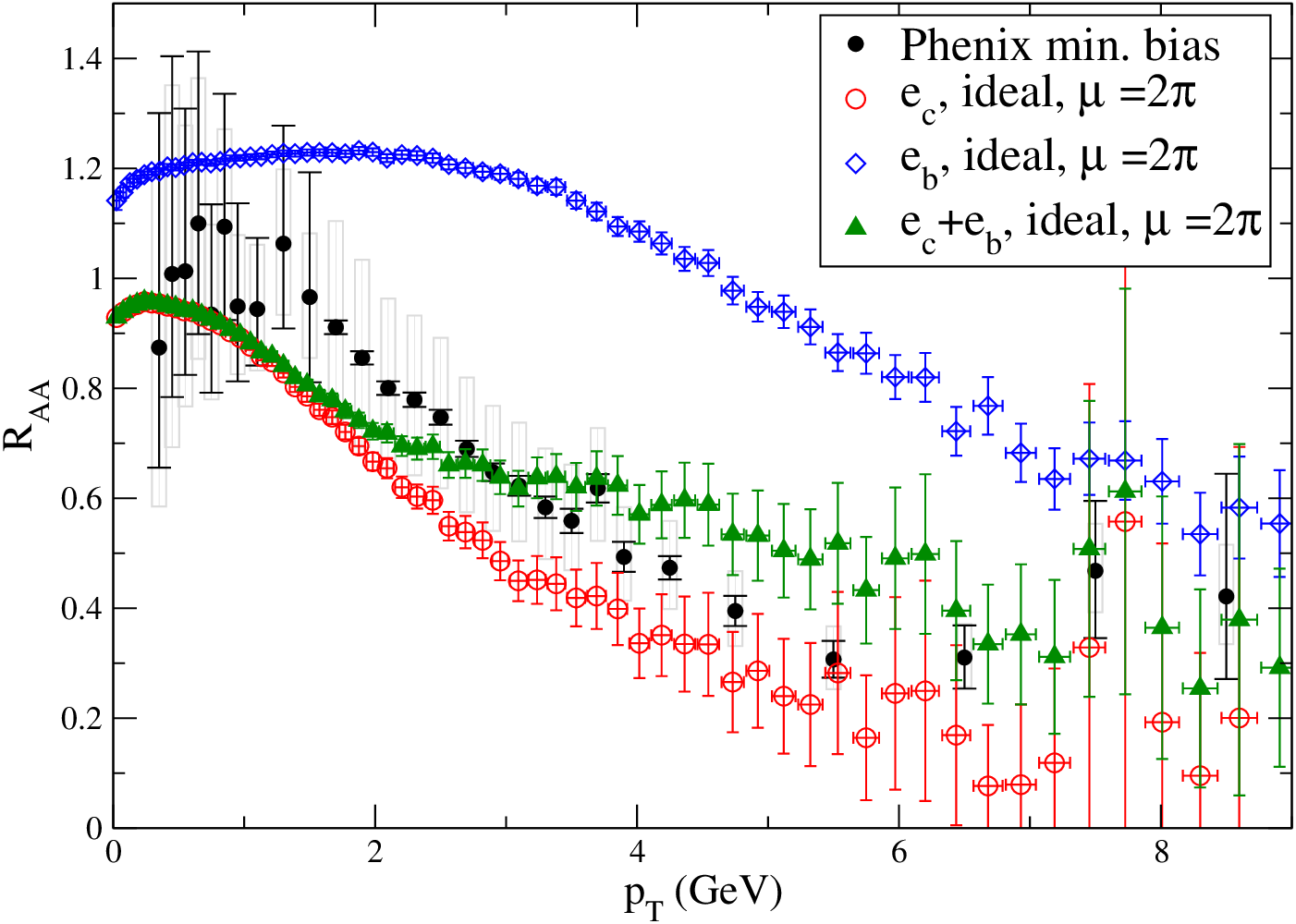}
\caption{Left panel: the effects of fragmentation and semileptonic decays, leading to a substantial quenching of the electron-$R_{AA}(p_T)$. Right panel: $R_{AA}(p_T)$ for the single-electron spectra from the decays of charm, bottom and their weighted combination. The latter can be compared with the experimental results by Phenix.} 
\label{fig:electrons}
\end{center}
\end{figure}
In the left panel of Fig.~\ref{fig:electrons} we consider the effects of fragmentation and decay into electrons: they both give rise to a quenching of the spectrum. Finally in the right panel we display, for a given scenario, the $R_{AA}(p_T)$ of the single-electron spectra and we compare it to the results by Phenix~\cite{Phenix}. Our results, obtained in a perturbative setup and referring to a quite moderate value of the coupling, appear in reasonable agreement with the experimental data. This also suggests that it should be worth addressing more carefully the role of collisional energy loss in the quenching of high-$p_T$ spectra, so far mainly attributed to gluon radiation.   

\end{document}